\documentclass[]{aastex631}

\usepackage{subfigure}
\usepackage{longtable}
\usepackage{threeparttable}
\usepackage{amsmath}

\begin{document}

\title{V455 Car: an oscillating eclipsing Algol-type binary in triple star system}

\correspondingauthor{Wen-Ping Liao}
\email{$^{*}$liaowp@ynao.ac.cn}	
\author{Zhao-Long Deng}
\affiliation{Yunnan Observatories, Chinese Academy of Sciences (CAS), 650216 Kunming, China}
\affiliation{University of Chinese Academy of Sciences, No.1 Yanqihu East Rd, Huairou District, Beijing, China 101408}

\author{Wen-Ping Liao$^{*}$}
\affiliation{Yunnan Observatories, Chinese Academy of Sciences (CAS), 650216 Kunming, China}
\affiliation{University of Chinese Academy of Sciences, No.1 Yanqihu East Rd, Huairou District, Beijing, China 101408}

\author{LiYing Zhu}
\affiliation{Yunnan Observatories, Chinese Academy of Sciences (CAS), 650216 Kunming, China}
\affiliation{University of Chinese Academy of Sciences, No.1 Yanqihu East Rd, Huairou District, Beijing, China 101408}

\author{Xiang-Dong Shi}
\affiliation{Yunnan Observatories, Chinese Academy of Sciences (CAS), 650216 Kunming, China}

\author{Nian-Ping Liu}
\affiliation{Yunnan Observatories, Chinese Academy of Sciences (CAS), 650216 Kunming, China}

\author{Ping Li}
\affiliation{Yunnan Observatories, Chinese Academy of Sciences (CAS), 650216 Kunming, China}



\begin{abstract}

 V455 Car is a southern oscillating eclipsing Algol-type system with an orbital period of 5.132888 days. Our first photometric solutions based on the Transiting Exoplanet Survey Satellite indicate that it is a semi-detached binary with the secondary star is almost filling its Roche lobe. The noticeable O'Connell effect in light curve could be explained by hot spot on the primary component, which may be attributed to the mass transfer from the secondary component to the primary one.  The absolute parameters are determined as: $M_{1} = 5.30 \pm 1.10 \, \rm M_{\odot}$, $R_{1} = 3.17 \pm 0.22 \, \rm R_{\odot}$ for the primary, and $M_{2} =  1.58 \pm 0.32 \, \rm M_{\odot}$, $R_{2} = 6.66 \pm 0.46 \, \rm R_{\odot}$ for the secondary. Based on $O-C$ analysis, we find a periodic variation of $P_3=26.62(\pm1.66)\,yr$. The periodic oscillation suggests a possible third body with a minimal mass of $0.59(\pm0.13)\,\rm M_{\odot}$. It is speculated that the secondary star has undergone a longer evolution, leading to a mass ratio reversal being experienced in the binary system. Our frequency analysis finds that the primary of V455 Car may be an SPB/SLF star. This study reports a novel example of an oscillating eclipsing Algol-type system featuring an SPB/SLF primary star and a red giant star, which suggest that strong observational results for a high incidence of third bodies in massive binaries.
\end{abstract}



\section{Introduction}
\label{introduction}

Slowly Pulsating B stars (SPB stars) and Beta Cephei variables (BCEP stars) are the only two recognized types of OB-type pulsating variables in the upper main sequence \citep{shi2023observational}. SPB stars, typically of late B-type spectral type (approximately B3 to B9) with masses ranging from 2.5 to 8 $\rm M_{\odot}$, exhibit non-radial multi-periodic g-mode pulsations with periods of 0.5 to 3 days \citep{2010aste.book.....A} driven by $\kappa$-mechanism. In contrast, BCEP stars are characterized by p-mode pulsations driven by the same $\kappa$-mechanism, activated in the ionization zones of iron-group elements. SPB stars were first categorized by \citet{1991A&A...246..453W}, and BCEP stars have been extensively cataloged, with recent expansions from ground-based and space-based surveys \citep{Stankov_2005, 2005AcA....55..219P, 2019MNRAS.489.1304B, 2020AJ....160...32L}. Additionally, some B-type stars exhibit stochastic low-frequency (SLF) variability, characterized by quasi-periodic, time-dependent oscillations spanning minutes to days, attributed to internal gravity waves (IGWs) excited at convective core boundaries or turbulent envelopes \citep{2019A&A...621A.135B, 2020A&A...640A..36B}. Space-based missions such as CoRoT, Kepler, K2, and TESS have significantly enhanced our understanding of the internal processes in massive stars.\citep{aliccavucs2022analysis, gilliland2010kepler, howell2014k2, ricker2015transiting}.

V455 Car is an oscillating eclipsing Algol-type (oEA) binary system with an orbital period of 5.132888 days from VSX \footnote{\url{https://www.aavso.org/vsx/index.php?view=detail.top&oid=6208}}. However, the pulsating characteristic of V455 Car is still unknown. It was firstly identified as an EA-type binary star by \citet{1999IBVS.4659....1K}, which was listed in the TESS Eclipsing Binary Catalog \citep{2022ApJS..258...16P}, and its spectral type was preliminarily identified as B5/B6 V by \citet{1975mcts.book.....H}. The effective temperature of 18000 K was estimated by \citet{2013ApJS..208....9P}. \citet{tokovinin2018updated} made preliminary estimation of the angular separation between two component stars (0.157 mas) and mass of the primary star (3.49 $ \rm M_{\odot}$). \citet{2019ApJ...872L...9P} classified V455 Car as a B-type pulsating eclipsing binary system through classification of TESS light curves. Subsequently, \citet{2022ApJS..258...16P} gave an EB-type binary preliminary classification and epoch of V455 Car. Its characteristics of pulsating component co-existence in binary system and unknown pulsation features make it an ideal object for our research.

This paper is organized as follows, Section 2 describes the data acquisition and processing of TESS observations and FEROS spectroscopic data. We derived the atmospheric parameters of the primary star.
Section 3 presents the first photometric solutions based on the TESS light curves. In Section 4, we analyze the orbital period variation of V455 Car. The $O-C$ curves were described by the light travel-time effect (LTTE) due to the third body. Section 5 displays the pulsation analysis, indicating that V455 Car may be an SPB/SLF star. In the last Section 6, we discuss the evolution state, spot activity, and the third body.

\section{Data acquisition}
%%\label{}
\subsection{Photometric data}
\begin{figure*}
    \centering
    \includegraphics[width=0.5\textwidth, angle=0]{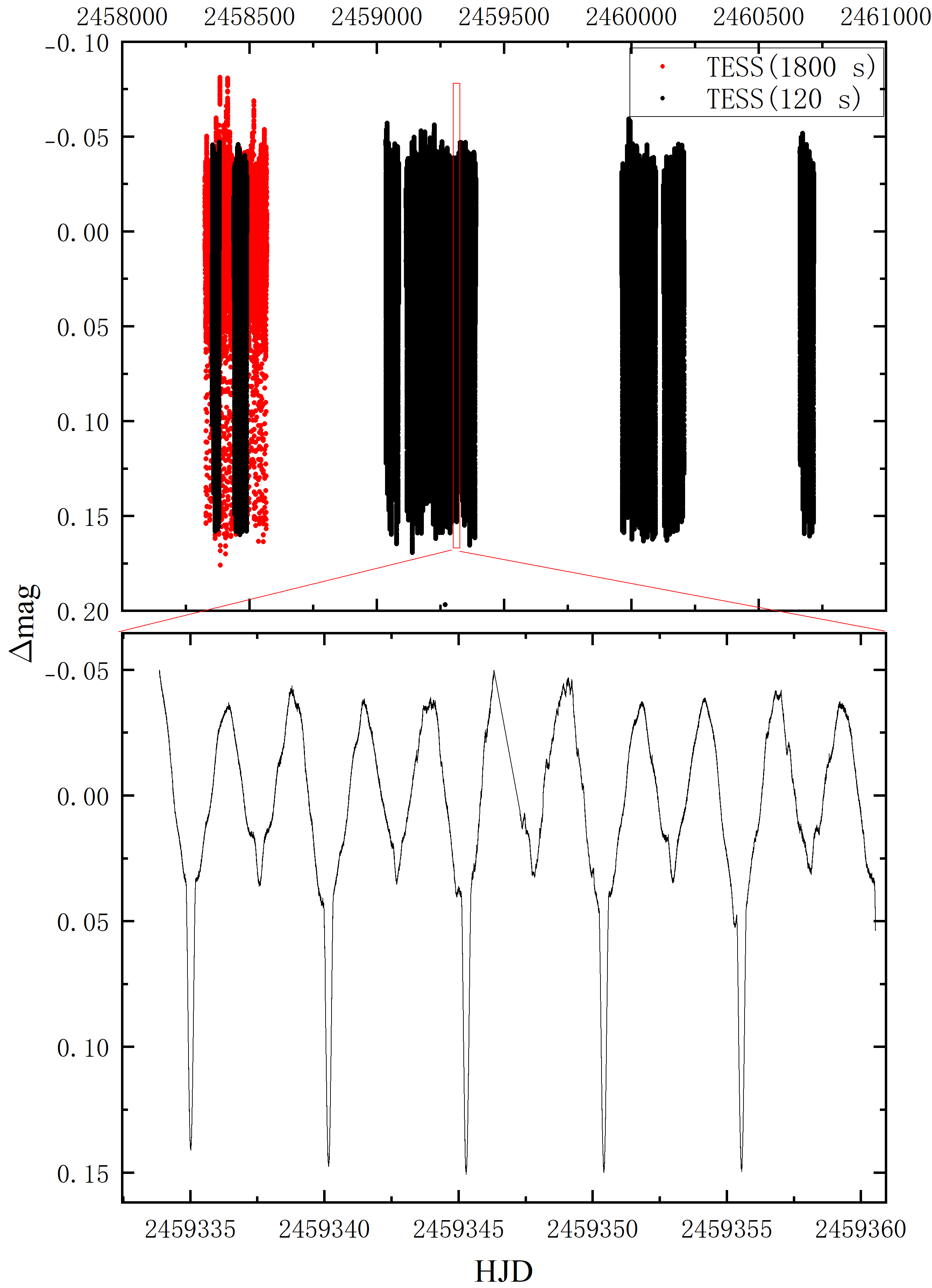}
    \caption{TESS light curves of V455 Car. The upper panel: all light curves after detrending processing from TESS, 120 s-cadence and 1800 s-cadence light curves are shown as black and red points, respectively. The lower panel: light curve segments for photometric analysis.\label{fig:figure1}}
  \label{fig:LC}
\end{figure*}

We download all available TESS data using the Mikulski Archive for Space Telescopes (MAST) \footnote{MAST: \href{https://archive.stsci.edu/mast.htm}{$archive.stsci.edu$}} database. All of the observations from HJD 2458325 to 2460718 were obtained in Sectors 2, 5, 6, 27-28, 30-39, 61-69, 87 and 88, including short exposure cadence data (120 s), medium exposure cadence data (600 s) and long exposure cadence data (1800 s) in Sectors 1, 3, 4, 7-9. The photometric band range for TESS is 600-1000 nm to achieve a photometric accuracy of 50 ppm on stars \citep{ricker2015transiting}.  Based on the comparative study between SAP and PDCSAP light curves \citep{10.1093/mnras/stae1352}, we choose to convert SAP-flux to magnitude, and the time was converted to Heliocentric Julian Day (HJD), then the TESS data were phased out using the method recommended by \cite{Zhang_2020}. Figure \ref{fig:figure1} shows TESS light curves of V455 Car, where the upper panel displays 120 s-cadence and 1800 s-cadence light curves after detrending processing from TESS and the lower panel presents light curve segments for photometric analysis. We also collected photometric data from the All-Sky Automated Survey (ASAS) \citep{2002AcA....52..397P} and All-Sky Automated Survey for Supernovae (ASAS-SN) \citep{shappee2014man,kochanek2017all,2018MNRAS.477.3145J} catalogues, the ASAS-SN raw data were processed with the machine learning saturated star photometry because raw flux reached saturation \citep{winecki2024photometry}. This method uses a multilevel perceptron (MLP) neural network to obtain photometry of saturated stars, which provides significantly better results for saturated stars.

\subsection{FEROS spectrum}
\begin{figure}
  \centering
  \includegraphics[width=0.5\textwidth, angle=0]{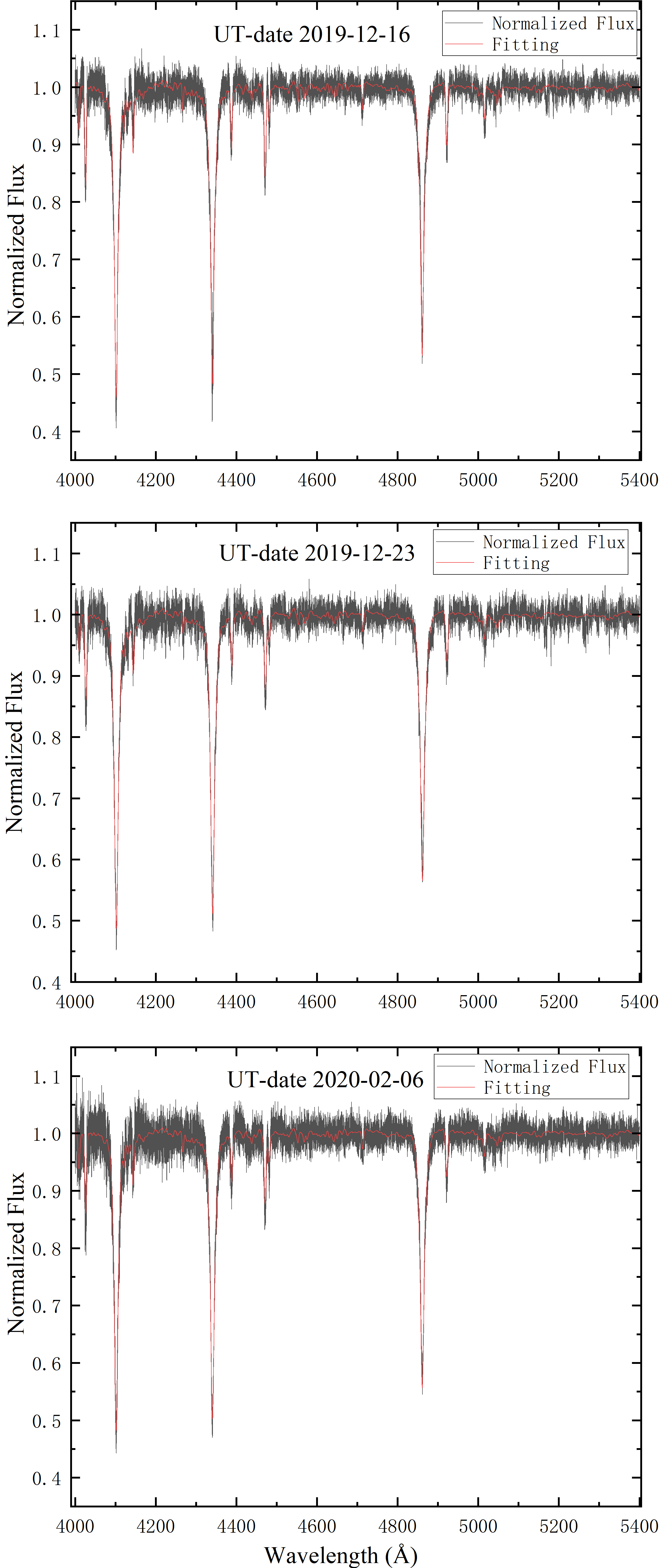}
  \caption{Fitting results for all FEROS normalized spectra. The normalized spectral data and fitting results are displayed as black and red lines, respectively.\label{fig:figure2}}
  \label{fig:spectra-fit}
\end{figure}

Utilizing data from Gaia DR2, Gaia DR3 and \citet{2013ApJS..208....9P}, the effective temperatures are estimated to 9386.50$^{+309.50}_{-1241.75}$ K, 17661.94$^{+261.99}_{-233.40}$ K, and $18000$ K by them, respectively, the errors of Gaia temperature are from the Mikulski Archive for Space Telescopes (MAST). However, due to the unreliability of the Gaia GSP-Phot data \citep{10.1093/mnras/stad3601}, it is necessary to recalculate the effective temperature and other stellar atmosphere parameters. To address the limitations identified in the Gaia GSP-Phot data, we employed high-resolution spectra from the FEROS spectrograph to independently verify the spectral parameters of V455 Car. FEROS is a high-resolution spectrograph (R $\sim$ 48 000) covering a wavelength range of 3600–9200 Å \citep{1999Msngr..95....8K,2022A&A...665A..36G}. Considering that the temperature ratio $\frac{T_{2}}{T_{1}}$ of the primary and secondary star is small, we consider the primary's contribution to be dominant. Therefore, we used the ULySS (University of Lyon Spectroscopic Analysis Software, \citet{2009A&A...501.1269K}) to derive the atmospheric parameters, which is constructed by interpolating the MILES library \citep{2006MNRAS.371..703S,2011A&A...531A.165P}. All subsequent analyses are listed in Table \ref{tab:spec-Teff}, and the fitting spectra are listed in Figure \ref{fig:spectra-fit}. 
As can be seen from the Table \ref{tab:spec-Teff}, the orbital phases corresponding to 2019-12-16 and 2019-12-23 are close to 0.5, so the atmospheric parameters of the primary star are only taken from the mean values of
these two spectra: $T_{eff}$ =16427 $\pm$ 147 K, $\log g$ = 4.04 $\pm$ 0.04, $[Fe/H]$ = -0.09 $\pm$ 0.02. This temperature value should correspond to spectral type of $\sim$B4 V from the online table \footnote{\label{fn:myurl} \url{http://www.pas.rochester.edu/~emamajek/EEM_dwarf_UBVIJHK_colors_Teff.txt}} of \citet{2013ApJS..208....9P}.

\begin{table}
\caption{Derived atmospheric parameters for the V455 Car.}
\begin{center}
\setlength{\tabcolsep}{2.0mm}{
\begin{tabular}{lcccc}\hline\hline
UT-date	&	Phase	&	$T_{eff}$	&	$\log g$	&	$[Fe/H]$	\\
 	&	 	&	 (K)	&	($\rm cm/s^{2}$)	&	(dex)   \\
\hline
2019-12-16	&	0.357 	&	16261$\pm$133	&	4.05$\pm$0.03	&	-0.08$\pm$0.02	\\
2019-12-23	&	0.497 	&	16592$\pm$63	&	4.02$\pm$0.02	&	-0.10$\pm$0.01	\\
2020-02-06	&	0.714 	&	16295$\pm$90	&	3.93$\pm$0.03	&	-0.05$\pm$0.03	\\
\hline
\end{tabular}}
\end{center}
\label{tab:spec-Teff}
\end{table}

\section{Analysis of the TESS light curve}
\begin{figure*}
     \centering
     \includegraphics[width=0.6\textwidth,angle=0]{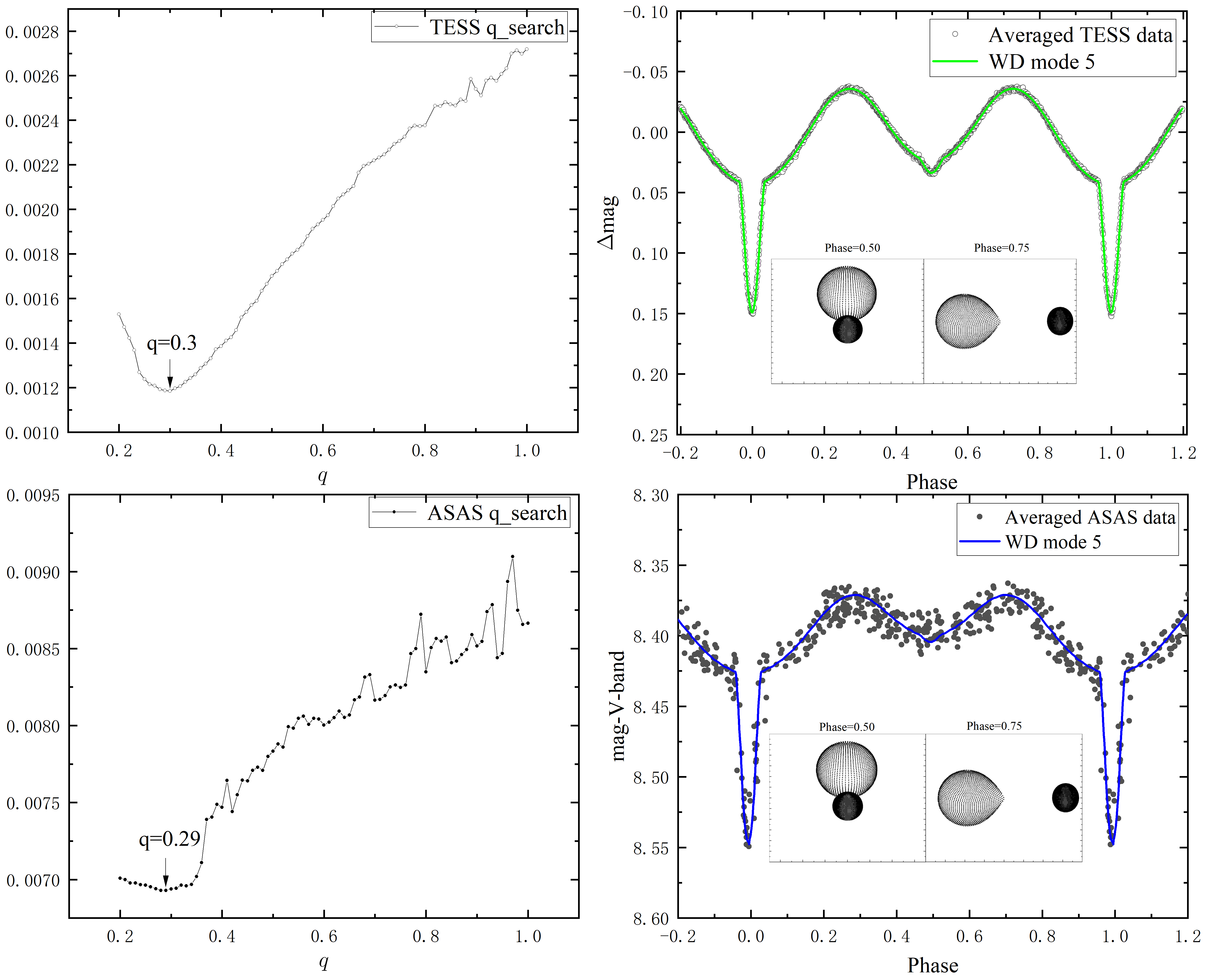}
     \caption{The upper left panel: the relationship between $q$ and the mean residuals $\overline{\Sigma}$ for W-D Mode 5 in TESS data. The upper right panel: the averaged TESS light curve (circles), the theoretical fit curve (green line for Mode 5), and the geometric structure. The lower left panel: the relationship between $q$ and the mean residuals $\overline{\Sigma}$ for W-D Mode 5 in ASAS data. The lower right panel: the averaged ASAS light curve (black dots), the theoretical fit curve (blue line for Mode 5), and the geometric structure.   \label{fig:figure 3}}
   \label{fig:LC-fit}
 \end{figure*}

In this section, we use the 120 s-cadence data of TESS and some data from ASAS \footnote{\url{https://www.astrouw.edu.pl/cgi-asas/asas_variable/073625-6152.4,asas3,5.132888,1907.7080,500,0,0}} to solve photometric solutions. As one can see from the upper panel of the Figure \ref{fig:figure1}, the light curves exhibit a fluctuating phenomenon, which could be attributed to the trend not being completely removed. To derive more accurate photometric solutions from light curves modelling, we implemented the following data processing steps: we selected a segment of symmetric light curve where the two maxima are almost equally high, which is shown in the lower panel of Figure \ref{fig:LC} (HJD 2459350-2459357) at sector 38, then phase-fold the time-domain light curve. Then, we removed the extracted pulsating components to minimize the effect of the pulsation on the light curve. Finally, we averaged the phase-folded light curves by averaging all data points within every 0.002 phase interval (those circles displayed in the upper right panel of Figure \ref{fig:LC-fit}). The processed light curve was then fitted using the Wilson-Devinney (W-D; \citealp{1971ApJ...166..605W,1979ApJ...234.1054W,1994PASP..106..921W}) code to obtain basic photometric solutions. 

During the process of modelling light curve with the W-D code, we chose Mode 5 (semi-detached binary with the secondary star filled with its Roche lobe) to fit. The primary effective temperature is fixed as $T_{eff1} = 16427$ K. It is evident from the upper right panel of Figure \ref{fig:LC-fit} that secondary eclipse is located at phase 0.5, allowing us to assume an orbital eccentricity of 0. Then we can set the bolometric albedo $A_1$ and $A_2$ to 1 and 0.5 \citep{rucinski1969proximity}, and set gravity-darkening coefficients $g_1$ and $g_2$ to 1 and 0.32, respectively \citep{1967ZA.....65...89L}. The free parameters
include the phase shift, the orbital inclination ($i$), the effective surface temperature of the secondary ($T_2$), the monochromatic luminosity of the primary ($L_1$), the modified dimensionless surface potential ($\Omega_1$ for Mode 5).

Firstly, we tried to solve the TESS data, and the mass ratio ($q = M_2/M_1$) can be determined using the q-search method. For Mode 5, the range of $q$ was set between 0.2 to 1 with a step of 0.01, and the minimal mean residual achieved at $q = 0.30$. Then we obtained the final basic photometric solutions through continuous iterations after setting $q$ as a free parameter. The mass ratio of V455 Car is determined to be $q = 0.298 \pm 0.002$ from Mode 5, and the primary star contributes most of luminosity to the total system (about 68 percent). To further confirm the results, we also performed photometric analysis on the ASAS data, the q-search was also set between 0.2 to 1 with a step of 0.01, and the minimal mean residual achieved at $q = 0.29$ in ASAS data, the mass ratio from ASAS data is determined to be $q = 0.285 \pm 0.009$ by similar steps. All of q-search results and the fitting effect are displayed in Figure \ref{fig:LC-fit}, where points and lines represent the averaged and the theoretical light curves for Mode 5, the geometric structure of V455 Car is also shown in the right panel. All photometric solutions are listed in the Table \ref{tab:table 2}. From the results, we can see that the luminosity ratio of $L_{1,2}/(L_{1} + L_{2})$ varies between the \emph{ASAS} and \emph{TESS} bands, this may result from variances in photometric band ranges between \emph{ASAS} and \emph{TESS}. The secondary star’s radius is about twice larger than that of the primary one, indicating that the secondary star may be a
giant. Finally, since the ASAS data is relatively diffuse, we chose the photometric solutions obtained from TESS for subsequent results.

\begin{table*}
	\begin{center}
		\footnotesize
		\caption{Photometric solutions of Mode 5 from TESS and ASAS photometry of V455 Car. The numbers in parentheses are the errors on the last bits of the data. $f_{1,2}$ represents the filling factors, which is the ratio of the stars' volume to their Roche lobe volume ($V_{star}$/$V_{RL}$).\label{tab:table 2}}
		\begin{tabular}{lcc}\hline\hline
Parameters & TESS & ASAS \\  
\hline
mode 5	&	 semi-detached	&	 semi-detached	\\   
$i$	(deg) &	70.25(3)&	70.52(19)\\  
$q=M_{2}/M_{1}$	&	0.2978(17)&	0.2848(87)\\  
$T_{1}$ (K)&    16427 (fixed)&  16427 (fixed)\\
$T_{2}$ (K)&	5619(12)&	5447(34)\\  
$L_{1}/(L_{1}+L_{2})_{BAND}$	&	0.6835(7)&	0.8362(14)\\   
$L_{2}/(L_{1}+L_{2})_{BAND}$	&	0.3165(7)&	0.1638(14)\\   
$g_{1}$	&	1.0	&	1.0	\\  
$g_{2}$	&	0.32&	0.32\\  
$A_{1}$	&	1.0	&	1.0	\\   
$A_{2}$	&	0.5&	0.5\\   
$\Omega_{1}$	&	7.82(3)&	7.97(14)\\   
$\Omega_{2}$	&	2.461(5)&	2.43(6)\\  
$r_{pole1}$	&	0.1329(5)&	0.1307(24)\\   
$r_{pole2}$	&	0.2605(12)&	0.2566(22)\\   
$r_{point1}$	&	0.1333(5)&	0.1310(24)\\   
$r_{point2}$	&	0.3783(4)&	0.3730(22)\\   
$r_{side1}$	&	0.1331(5)&	0.1308(24)\\ 
$r_{side2}$	&	0.2712(4)&	0.2671(23)\\   
$r_{back1}$	&	0.1333(3)&	0.1310(24)\\  
$r_{back2}$	&	0.3039(4)&	0.2998(23)\\   
$R_{2}/R_{1}$	&	2.102(5)&	2.121(23)\\   
$f_{1}$	&	2.053(14)&	1.882(64)\\  
$f_{2}$	&	99.2(6)&	99.2(30)\\  
mean residuals &  0.00108& 0.00558\\
\hline			
		\end{tabular}
	\end{center}
\end{table*}

\begin{figure*}
    \centering
    \includegraphics[width=0.7\textwidth, angle=0]{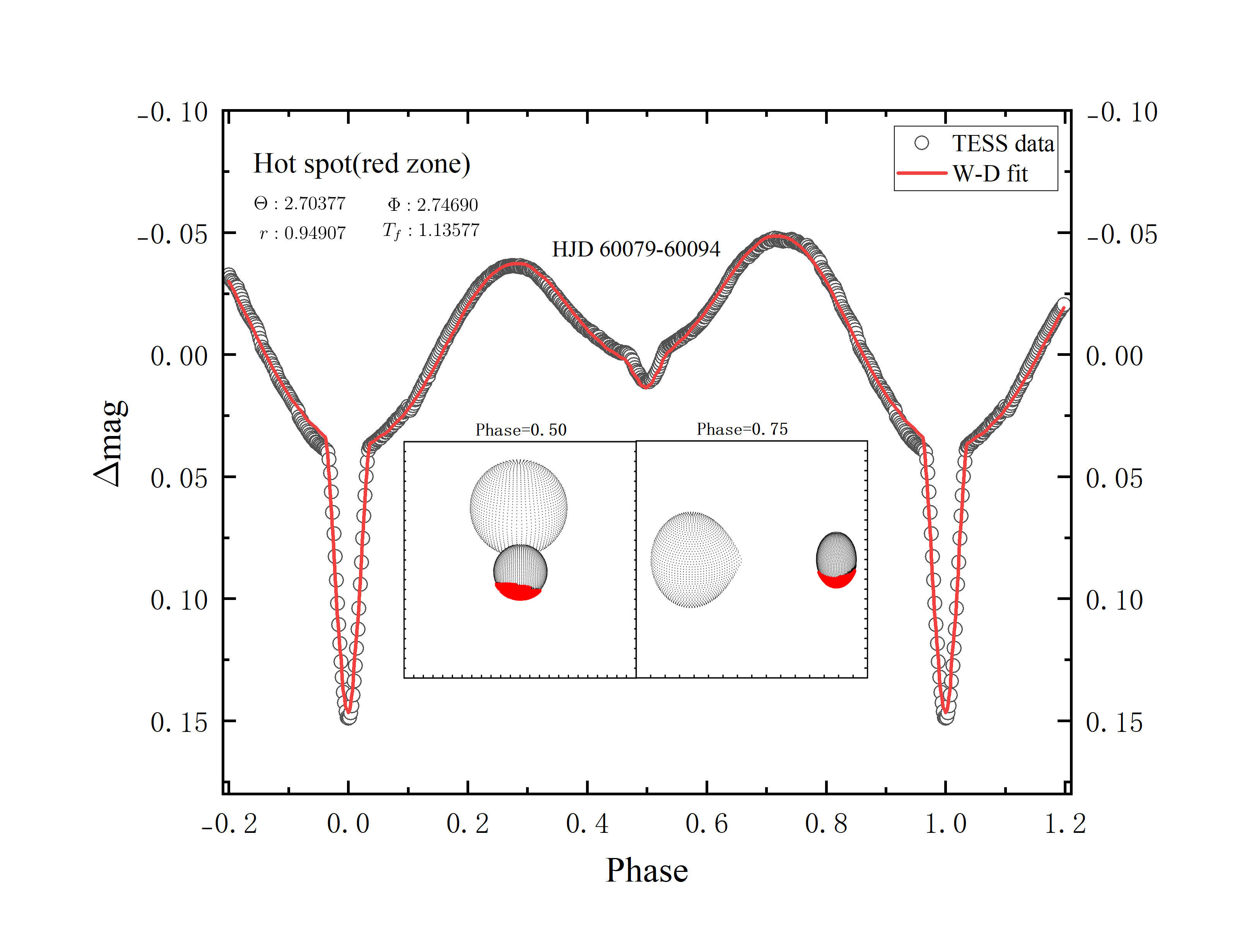}
 	\caption{Hot spot fitting obtained with the W-D program. The average asymmetric and theoretical light curves are displayed with open circles and red line, respectively. \label{fig:hotspot}}
 \end{figure*}

As shown in the upper panel of the Figure \ref{fig:LC}, the multi-period light curves from TESS display a visible O'Connell effect \citep{1969CoKon..65..457M}, which probably caused by one or more star spots on the surface of primary or secondary star. We consider the presence of hot spots on the primary component or dark spots on the secondary component as potential causes of the O'Connell effect. After adding free parameters of star spots, i.e., the co-latitude $\Theta$ (in radian), the longitude $\Phi$ (in radian), the radius $r$ (in radian), and the temperature factor $T_f$, we found the mean residuals value of the case that one hot spot on the primary component is smaller. The fitting results of hot spot solutions are shown in the Figure \ref{fig:hotspot},  and the spot parameters are also listed in this figure. In conclusion, the O'Connell effect of V455 Car is likely attributable to the hot spot on the surface of primary star. Considering the volume filling factors for the primary and secondary components are $2.053 (\pm 0.014)\%$ and $99.2 (\pm 0.6)\%$, the hot spot may be the result of mass transfer from the secondary component to the primary one.

\section{O-C analysis and light travel-time effect}

Analyzing the orbital period variations of binary star systems is also important. A commonly used method for studying orbital period changes is the O-C analysis (Observed minus Calculated analysis). In fact, the period variations of oEA systems are extremely useful, as they can provide information about mass transfer, magnetic activity, additional objects around the central binary and their potential impact on the pulsation behavior of the massive component \citep{2024arXiv240516365L}. We convert the data spanning multiple periods into phases and then use parabolic fits to obtain primary eclipsing times of TESS, Hipparcos, ASAS and ASAS-SN data \citep{2021AJ....161...46S,2022ApJ...924...30L}. Finally we get 159 primary eclipsing times, and all the primary eclipsing times of V455 Car are provided in a machine readable format in Table \ref{tab:primary}.

We use the following linear ephemeris for V455 Car provided by O-C gateway \footnote{{\url{https://var.astro.cz/en/Stars/10297}{O-C gateway (astro.cz)} }} to calculate $(O-C)$ values: 

\begin{eqnarray}\label{equation(2)}
	\mathrm{MinI} (\rm {HJD}) =2448504.5680 + 5^{\textrm{d}}.132888\times{E}
\end{eqnarray}

\begin{figure}
    \centering
    \includegraphics[width=0.6\textwidth, angle=0]{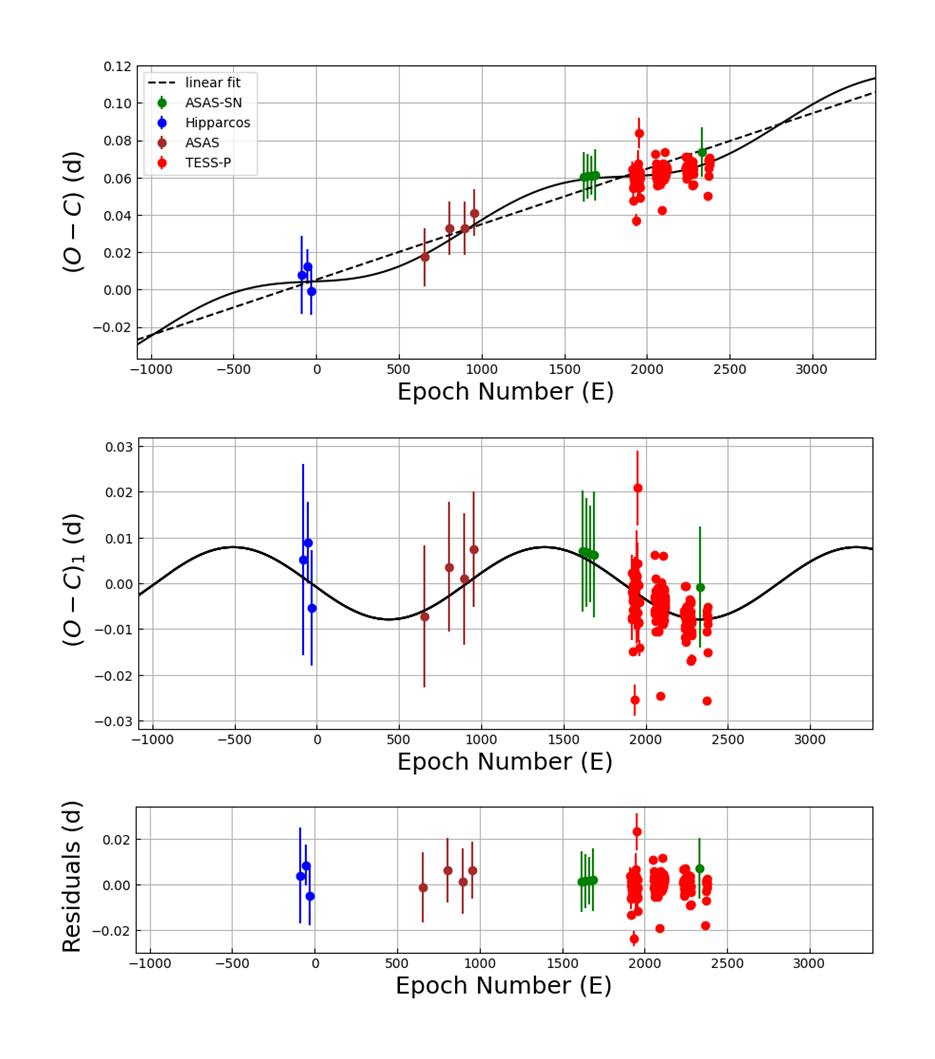}
 	\caption{Upper panel: $(O-C)$ diagram of V455 Car calculated with the linear ephemeris of Equation \ref{fit)}. The solid line refers to a combination of a linear ephemeris and a cyclic period variation, and the dashed line to a new linear ephemeris. Middle panel: $(O-C)_1$ curve after removing linear trend, the solid line
refers to the periodic oscillation fitting result. Lower panel: the residuals of whole fitting effect.\label{fig:O-C fit}}
 \end{figure}

As shown in the upper panel of Figure \ref{fig:O-C fit}, the distribution of the $(O-C)$ data shows a linear variation and periodic oscillation exists which may be caused by the light travel-time  effect (LTTE) of a third body \citep{liao2010most}. So we use the following equation with a circular orbit ($e_3=0$) assumption to fit the $O-C$ diagram:
\begin{eqnarray}\label{fit)}
\mathbf{(O-C)=\Delta T_0 + \Delta P \times{E} + A sin(\frac{2\pi}{P_3} E + \phi)}
\end{eqnarray}

After performing a fitting to the $(O-C)$ data, we can obtain the results as follows:
\begin{align}\label{equation(2)}
\text{\textbf{MinI}} (\mathrm{HJD}) &= 2458504.5780 \ (\pm 0.0034) \nonumber \\
&\quad + 5^{\mathrm{d}}.1329158 \ (\pm 0.0000016) \times E \nonumber \\
&\quad + 0.0079 \ (\pm 0.0020) \sin\left( \frac{2\pi}{1894.5\ (\pm 118.4)} E - 3.02 \ (\pm 0.33) \right)
\end{align}

$(O-C)_1$ curve after removing linear trend is displayed in the middle panel of Figure \ref{fig:O-C fit}, where the solid line
refers to the periodic oscillation fitting result. The  residuals of whole fitting effect are shown in the lower panel of Figure \ref{fig:O-C fit}, where $E$ represents the epoch of the primary minima in cycles. It is evident that the linear trend in the $(O-C)$ data has been effectively removed.  The semi-amplitude and period of the cyclical variation are $A=0.0079 (\pm0.0020)\,days$ and $P_3=26.62 (\pm1.66)\,yr$.

\begin{table*}
    \footnotesize
    \caption{ All the primary eclipsing times of V455 Car. This table is available in full in machine-readable form.\label{tab:primary}}
    \begin{tabular*}{\textwidth}{@{\extracolsep{\fill}}lll|lll|lll}\hline
HJD(Min I) & ERROR & SOURCE & HJD(Min I) & ERROR & SOURCE & HJD(Min I) & ERROR & SOURCE \\ \hline
2448063.14755 & 0.02091 & Hipparcos & 2459047.57962 & 0.00061 & TESS   & 2459386.35525 & 0.00060 & TESS   \\
2448232.53748 & 0.00888 & Hipparcos & 2459052.71630 & 0.00057 & TESS   & 2459966.37171 & 0.00056 & TESS   \\
2448350.58040 & 0.01264 & Hipparcos & 2459057.84883 & 0.00055 & TESS   & 2459971.50492 & 0.00053 & TESS   \\
2451871.76011 & 0.01534 & ASAS & 2459062.98346 & 0.00057 & TESS   & 2459976.63838 & 0.00057 & TESS   \\
2452641.70856 & 0.01423 & ASAS     & 2459068.11898 & 0.00062 & TESS   & 2459981.77661 & 0.00069 & TESS   \\
2453103.66848 & 0.01433 & ASAS      & 2459073.24268 & 0.00073 & TESS   & 2459986.90293 & 0.00073 & TESS   \\
.....&.....&.....&.....&.....&.....&.....&.....&..... \\
\hline
\end{tabular*}
\end{table*}

\section{Frequency analysis} \label{sec:Fourier}
Because the pulsation period range of SPB stars is generally between 0.5-3 days, identifying such pulsation frequencies requires longer continuous light curves. Any potential long-term trends could produce false signals in the low-frequency region, making the correct handling of these trends crucial for low-frequency analysis. After an experiment with different data and detrending methods, the PDCSAP FLUX of TESS SPOC data provide the best performance in removing low-frequency artifacts \citep{Ma_2024}. Therefore, we select the PDCSAP FLUX data (120 s) for further pulsation analysis, which were processed with the Pre-search Data Conditioning Pipeline \citep{jenkins2016tess}. For V455 Car, we selected the light curves of all sectors for pulsation analysis.

\begin{figure}
   \centering
    \includegraphics[width=1\textwidth, height=6.5cm,angle=0]{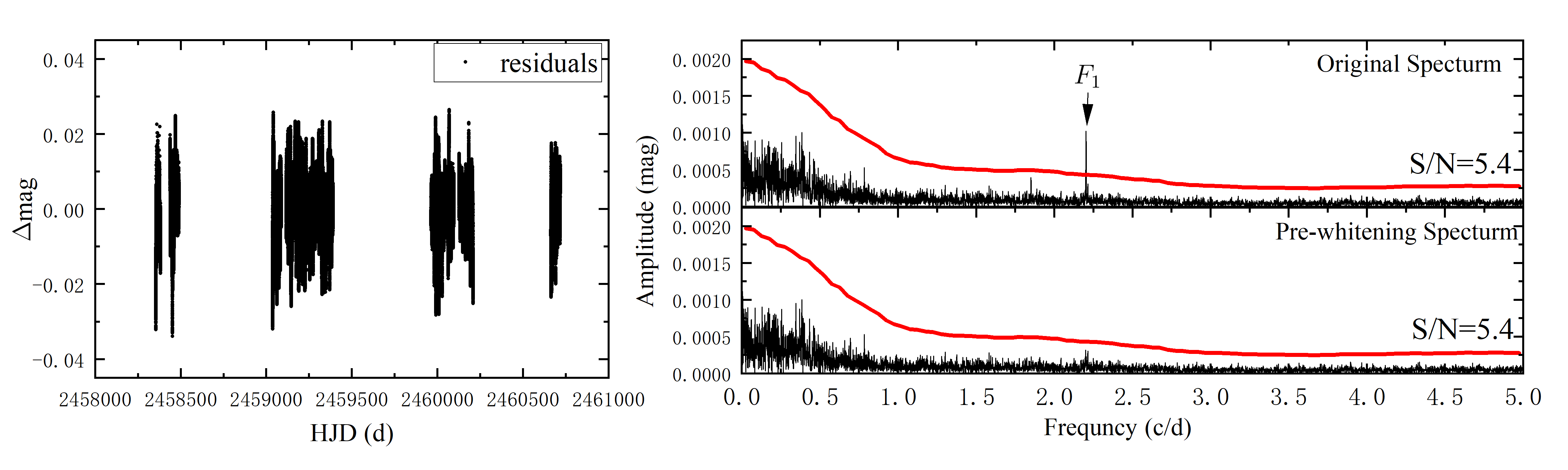}\caption{Left panel: residuals by removing eclipse light curves from the observed data and subtracting the twelve order Fourier series fit of the orbital frequency. Right upper panel: original spectrum obtained after residuals analysis. The frequency marked by arrow is extracted frequency, and the red line denotes the level of S/N = 5.4. Right lower panel: the spectrum of the residuals after pre-whitening. \label{fig:specturm}}
 \end{figure}

We obtained residuals and plotted it in the left panel of Figure \ref{fig:specturm} at by removing eclipse light curve from the observed data and subtracting the twelve order Fourier series fit of the orbital frequency (i.e. the multi-frequency harmonic mode of the orbital frequency; see e.g., \cite{yang2014photometric,2019ApJ...884..165Z,2021MNRAS.501L..65S}). To study the pulsation characteristics, we performed a multi-frequency analysis of the photometric residuals using the software Period04 \citep{2005CoAst.146...53L}. Since no pulsation signals were detected in the high-frequency region $(f > 5 d^{-1})$ and the pulsation periods of SPB-type stars are generally in the low-frequency region, we extracted frequencies only in the range of  $0-5   \ d^{-1}$. 

From the amplitude spectrum of the TESS light curve, we also used the Equation \ref{equation(5)} to pre-whiten the data:
\begin{eqnarray}\label{equation(5)}
	m(t)=m_0+\sum_{i=1}^{N}A_{i}\sin\left(2\pi\left(\nu_{i} (t\ - t_{0}) +\phi_{i}\right)\right)
\end{eqnarray}
where $\nu_i$ is the frequency extracted from the spectrogram, $A_i$ is the amplitude, $\phi_i$ is the phase. Each time a frequency is extracted, its corresponding sine function term is subtracted until the signal in the fourier amplitude spectrum is lower than the signal-to-noise ratio threshold, and the periodic component in the light curve can be considered to have been extracted.

We extracted frequencies following the rule of signal-to-noise ratio $S/N > 5.4$ \citep{2018aa,2021ApJ...920...76C,shi2023observational}. To identify independent and combination frequencies, we studied the linear combination of the pulsation frequency with the largest amplitude and the orbital frequency to match other frequencies. If the absolute difference between other frequencies and the linear combination is less than $\delta f$=1.5/$\Delta T$ ($\Delta T$ is the time span of the light curve), then the frequency is considered to a combination frequency. This process is iterated until all frequencies have been identified. The remaining frequencies are then considered as independent frequencies \citep{1978Ap&SS..56..285L,kurtz2015unifying}. We found one independent frequency with $S/N > 5.4$ fall within the pulsation period range of SPB stars. Its frequency parameters are  Freq. (d$^{-1}$) = 2.20216 (35), Ampl.(mag) = 0.001126 (136), Phase = 0.5903(13), and S/N = 14.87. The spectrum before and after pre-whitening are shown in the right panel of Figure \ref{fig:specturm}. We calculated the noise with a box size of 1 d$^{-1}$ and the original residuals \citep{shi2023observational}, the uncertainties of frequencies, amplitudes and phase are calculated using Monte Carlo simulations as described in \citet{fu2013asteroseismology}.

In conclusion, we can detect independent frequency in specturm, and we can also see apparent red noise from the residual amplitude spectrum in the right panel of Figure \ref{fig:specturm}, this means that the light curve also has SLF variability, which has been inferred to be caused by stochastically excited gravity waves driven by turbulent (core) convection \citep{2019A&A...621A.135B,2019NatAs...3..760B,2020A&A...640A..36B,2024A&A...692A..49B} or turbulence caused by sub-surface convection zones \citep{schultz2022stochastic}, which are an efficient mixing mechanism inside massive stars. So we currently suggested that the primary component may be an SPB or/and SLF. Further observations and analysis are needed to refine this conclusion.

\section{Discussions and conclusions}
%%\label{}

\begin{figure}
	 \centering
      \includegraphics[width=0.5\textwidth, angle=0]{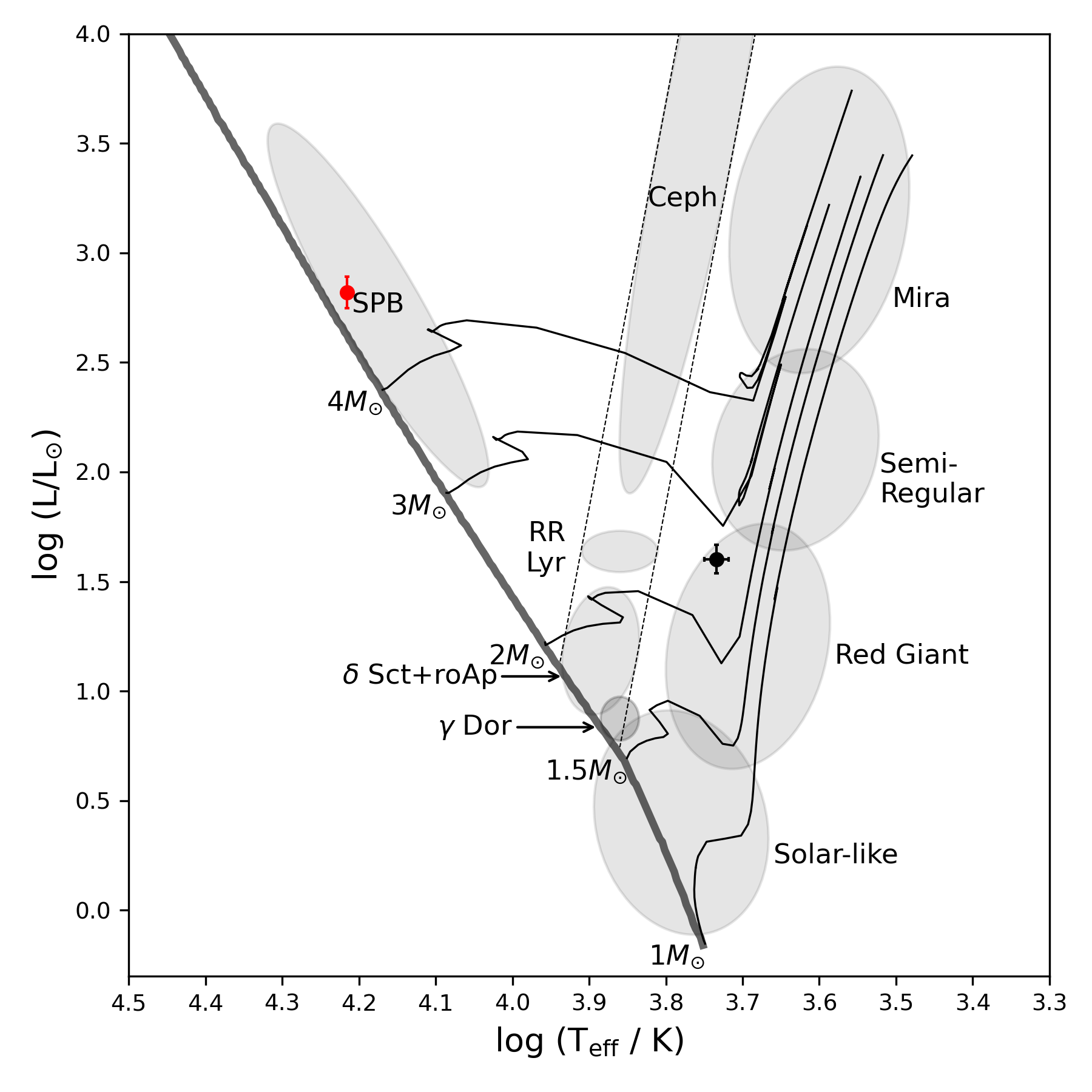}
 	\caption{The positions of the primary (red point) and secondary (black point) components on the H-R diagram \citep{2010aste.book.....A}. \label{fig:HR}}
\end{figure}

\begin{figure*}
	 \includegraphics[width=\linewidth]{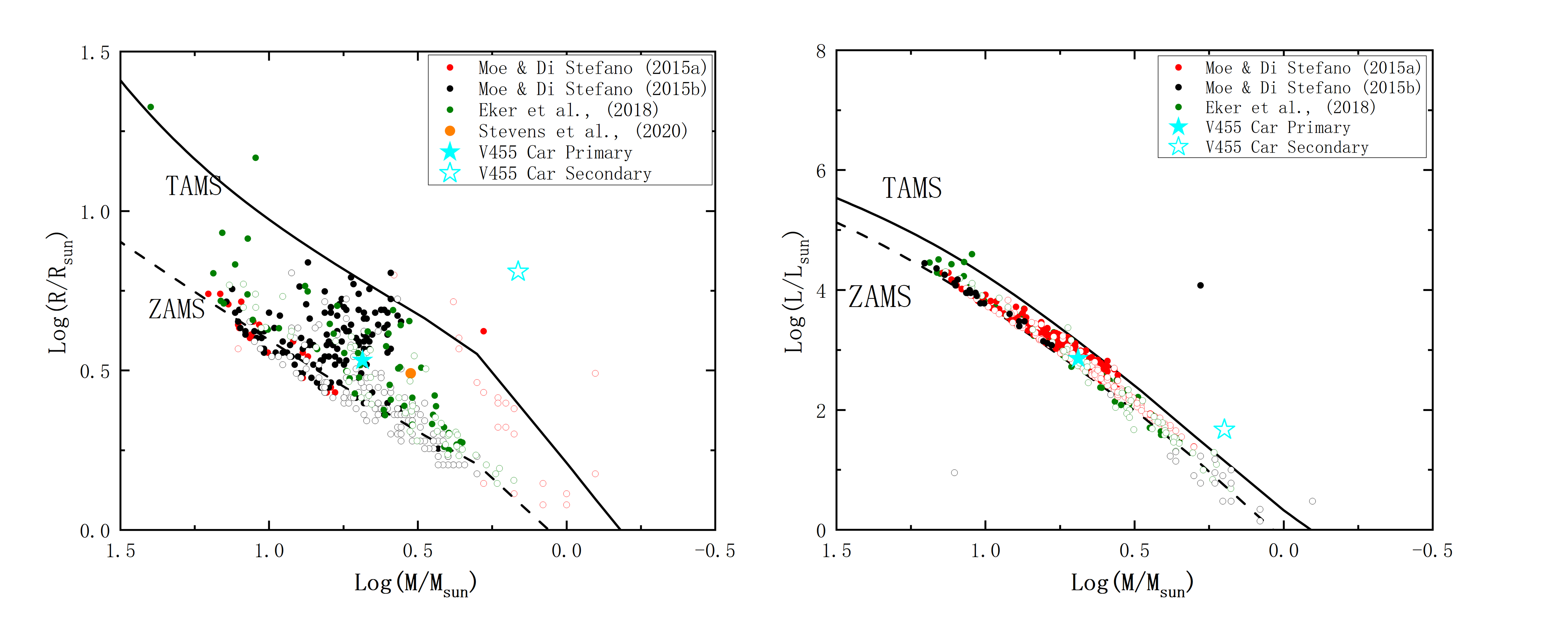}
\caption{Positions of V455 Car on mass-radius (M-R, left panel), the mass-luminosity (M-L, right panel). Solid symbols present the primaries, while the same symbols of hollow denote the low-mass components. Also displayed in the panels are the samples from \citet{2015ApJ...801..113M,2015ApJ...810...61M}, \citet{2018MNRAS.479.5491E}, \citet{2020MNRAS.499.3775S}.}
\label{HR}
\end{figure*}

We used W-D code for the first photometric analysis of TESS and ASAS data, it is found that V455 Car is a semi-detached binary with mass ratio $q = 0.298 \pm 0.002$ where the secondary star almost fills its Roche lobe. The noticeable O'Connell effect in light curve is attributed to the presence of hot spot on the surface of primary star, which may be caused by the mass transfer from the secondary star to the primary one. Additionally, the O'Connell effect may be a part of pulsation can not be excluded.

We collected and listed the values of $E(B-V)$, $m_B$ and $m_V$ from the ASAS-SN Catalog of Variable Stars: II (Jayasinghe et al., 2019) in Table \ref{tab:absolute parameters}, therefore $(B-V)_0 = (m_B- m_V)- E(B-V)=-0.172$. By considering that the primary component is a main-sequence (MS) star, we can use linear interpolation with $(B-V)_0$ to estimate the mass of primary star through the online table \footref{fn:myurl} of \citet{2013ApJS..208....9P}. So combing the photometric solutions, the masses of primary and secondary components are estimated to be $M_{1}$ = 5.30 $\pm$ 1.10 $\,\mathrm{M}_{\odot}$ and $M_{2}$ = 1.58 $\pm$ 0.32 $\,\mathrm{M}_{\odot}$ through $(B-V)_0$, where the uncertainty of the primary mass is 20\% of its mass \citep{li2024evolutionary}. Subsequently, by using the Kepler’s Third Law, the semi-major axis is calculated to be $a$ = 23.81 $\pm$ 1.56 $\,\mathrm{R}_{\odot}$, so we can calculated the mean radius for both components by combining the relative radius determined from the W-D with the semi-major axis as $R_{1} = 3.17 \pm 0.22 \,\mathrm{R}_{\odot}$ and $R_{2} = 6.66\pm 0.46 \,\mathrm{R}_{\odot}$. Their luminosities are estimated as $\log (L_1/ \mathrm{L}_{\odot})$ = 2.82 $\pm$ 0.06, and $\log (L_2/ \mathrm{L}_{\odot})$ = 1.60$\pm$ 0.07, from the $L=4 \pi R^{2}\sigma T_{\mathrm{eff}}^{4}$ with their relative temperature ratio and radius ratio modeled by the W-D program. All the absolute parameters are listed in Table \ref{tab:absolute parameters}. From the positions of the primary and secondary components on the H-R diagram (Figure \ref{fig:HR}), it is found that the primary star is in SPB instability zone, secondary star is located in the zone of red giant.

From the frequency-amplitude diagram in right panel of Figure \ref{fig:specturm}, we identified one frequency from the spectrum. Simultaneously, obvious SLF variability is found in the residual amplitude spectrum, which is common in masssive stars \citep{2019NatAs...3..760B,2020A&A...640A..36B,bowman2023making}. Currently, there are several explanations for the mechanism of SLF variability, such as Internal Gravity Waves (IGWs) or the dynamics of their turbulent envelopes \citep{schultz2022stochastic,2024A&A...692A..49B}. Previous studies on SLF variability have primarily focused on single stars, therefore the continued study of SLF variability in binary systems holds an exciting prospect.

Based on all available primary minima times, a linear trend and periodic oscillation were identified in the $O-C$ diagram. The periodic oscillation can be explained as the presence of a third body, the period of the third body was $P_{3}= 26.62 \pm 1.66$ yr. In order to have a further study on the third body, we can use the mass function to estimate parameters of the third body by setting the orbital inclination in $i_3=90^{\circ}$:

\begin{equation}
  f(m)=\frac{(M_3sini_3)^3}{(M_1+M_2+M_3)^2}=\frac{4\pi^2}{GP_3^2}\times (cA)^3,
  \label{eq:mass function}
\end{equation}

Then the mass function of the third body is $f(m)=0.0036(\pm0.0028)\,M_{\odot}$, the minimum mass of the third body is $M_3=0.59(\pm0.13)\,M_{\odot}$, and the largest projected semi-major axis of the hypothetical triple orbit is calculated to $16.06\pm5.91$ au. We also tried to add the third light in the photometric solution to find the evidence that the third body is a normal star. However, the resulting of $l_3$ is always close to zero, we speculate this third body may be a late K-type main sequence star which is much fainter than the central binary system.

\begin{table}[htbp]
  \centering
  \caption{Estimated absolute parameters of V455 Car.}
  \label{tab:absolute parameters}
  \begin{tabular}{@{}llll@{}}
    \hline
    Parameters         & Value           & Parameters        & Value           \\
    \hline
    $\varpi$ (mas)     & 1.8854(0.0351)  & $m_B$ (mag)            & 8.38            \\
    $m_V$ (mag)             & 8.378           & $E(B-V)$ (mag)         & 0.174           \\
    $M_1~(\mathrm{M}_{\odot})$ & 5.30(1.10) & $M_2~(\mathrm{M}_{\odot})$ & 1.58(0.32) \\
    $R_1~(\mathrm{R}_{\odot})$ & 3.17(0.22) & $R_2~(\mathrm{R}_{\odot})$ & 6.66(0.46) \\
    $L_1~(\mathrm{L}_{\odot})$ & 659(94)   & $L_2~(\mathrm{L}_{\odot})$ & 40(6)       \\
    Semi-major axis~$(\mathrm{R}_{\odot})$ & 23.81(1.56) & & \\
    \hline
  \end{tabular}
\end{table}
 
 The mass-radius and mass-luminosity positions are shown in Figure \ref{HR}, which has collected the samples from \citet{2018MNRAS.479.5491E}, \citet{2020MNRAS.499.3775S}. Also shown in the panels are 152 B-type binary stars in Large Magellanic Cloud (LMC) that were investigated by \citet{2015ApJ...801..113M,2015ApJ...810...61M}. We can see from the mass-radius diagram that the primary star is between ZAMS and TAMS, while the secondary star has passed through TAMS. This behavior may be due to the interactions between the two components. And the secondary star has undergone a longer evolution, it suggests that it was once the more massive primary star, so this star may expand its outer layers as it evolves, ultimately filling its Roche lobe and consequently transferring mass to its less massive companion, finally reversing the mass ratio \citep{1955ApJ...121...71C,1998A&AT...15..357P}.

In conclusion, V455 Car represents an example of a binary system with an SPB/SLF primary star. Because the secondary star of V455 Car almost fills its Roche lobe, but the filling factors of the primary star is extremely low, we speculate that it is an Algol-type system that has just experienced a rapid mass transfer stage \citep{wagg2024asteroseismic}, the periodic variation may be attributed to the presence of a third body, and the specific mass transfer mechanism and process can be further studied in future. This system provides a relevant case for study of the evolution of massive binary stars.

\section*{Acknowledgements}

This work is supported by  the International Cooperation Projects of the National Key R/\&D Program of China (No. 2022YFE0116800), the Science Foundation of Yunnan Province (No. 202401AS070046), and the International Partership Program of Chinese Academy of Sciences (No. 020GJHZ2023030GC), the 2022 CAS ``Light of West China" Program, the Young Talent Project of  ``Yunnan Revitalization Talent Support Program" in Yunnan Province, the basic research project of Yunnan Province (Grant No. 202201AT070092).
This work has made use of data from the European Space Agency (ESA) mission Gaia. Processed by the Gaia Data Processing and Analysis Consortium.Funding for the DPAC has been provided by national institutions, in particular the institutions participating in the $Gaia$ Multilateral Agreement. FEROS specturm is based on data obtained from the ESO Science Archive Facility with DOI(s): \href{https://doi.org/10.18727/archive/24}. The TESS data presented in this paper were obtained from the Mikulski Archive for Space Telescopes (MAST) at the Space Telescope Science Institute (STScI). STScI is operated by the Association of Universities for Research in Astronomy, Inc. Support to MAST for these data is provided by the NASA Office of Space Science. Funding for the TESS mission is provided by the NASA Explorer Program.


\clearpage\newpage

\bibliography{Ref}{}
\bibliographystyle{aasjournal}



\end{document}